\documentclass[aps,pre,twocolumn,superscriptaddress]{revtex4-1}
\usepackage{xcolor}
\usepackage{graphicx}
\usepackage{amsmath}
\usepackage{amsfonts}

\begin{document}
	
	
 	\title{Population heterogeneity in the fractional master equation, ensemble self-reinforcement and strong memory effects}
	
	
	
	\author{Sergei Fedotov}
	\email[]{sergei.fedotov@manchester.ac.uk}
	\affiliation{Department of Mathematics, University of Manchester, M13 9PL, UK}
	
	\author{Daniel Han}
	\email[]{dhan@mrc-lmb.cam.ac.uk}
	\affiliation{Medical Research Council, Laboratory of Molecular Biology, Neurobiology Division, Cambridge, UK}
	
	
	\date{\today}
	
	\begin{abstract}
		We formulate a fractional master equation in continuous time with random transition probabilities across the population of random walkers such that the effective underlying random walk exhibits \textit{ensemble self-reinforcement}. The population heterogeneity generates a random walk with conditional transition probabilities that increase with the number of steps taken previously (self-reinforcement). Through this, we establish the connection between random walks with a heterogeneous ensemble and those with strong memory where the transition probability depends on the entire history of steps. We find the ensemble averaged solution of the fractional master equation through subordination involving the fractional Poisson process counting the number of steps at a given time and the underlying discrete random walk with self-reinforcement. We also find the exact solution for the variance which exhibits superdiffusion even as the fractional exponent tends to 1.
		
	\end{abstract}
	
	
	\maketitle
	\section{Introduction}  Anomalous diffusion appears in many natural processes in physics, chemistry and biology when measurements of mean squared displacement $m^{(2)}(t)$ show non-linear dependence on time: $m^{(2)}(t)\propto t^{\mu}$ \cite{bouchaud1990anomalous,metzler1999anomalous,metzler2004restaurant,scalas2006application,henry2010fractional}. A variety of models has been suggested for anomalous diffusion including continuous time random walk \cite{klafter2011first}, fractional Brownian motion \cite{mandelbrot1968fractional}, generalized Langevin equation \cite{kubo1966fluctuation,zwanzig1973nonlinear,goychuk2007anomalous} and L\'evy walks \cite{shlesinger1987levy,zaburdaev2015levy}. A typical feature of anomalous transport models involving temporal subdiffusion and superdiffusion is the appearance of memory effects. When a stochastic process depends on a series of previous events, it is often referred to as having non-Markovian characteristics or memory. In many natural phenomena, memory is a recurring theme, such as earthquakes \cite{livina2005memory}, quantum physics \cite{piilo2008non,wolf2008assessing,luchnikov2020machine} intracellular transport \cite{bressloff2013stochastic,chen2015memoryless,ba2018whole,fedotov2018memory}, and cell motility \cite{huda2018levy}.
Another direction to model anomalous diffusion is through random walks that account for the whole history of its past, described as strong memory \cite{schutz2004elephants,kumar2010memory,da2013non,budini2016inhomogeneous,baur2016elephant,pemantle1988phase}. However, it is difficult to justify why natural processes should exhibit such strong memory effects as seen in elephant random walks \cite{schutz2004elephants}, especially for inanimate objects such as intracellular organelles. In efficient search strategies \cite{tejedor2012optimizing} that have an essential role in time sensitive biological processes \cite{lanoiselee2018diffusion}, strong memory has significant effcts \cite{meyer2021optimal}. More recently, it was shown that strong memory and reinforcement can generate superdiffusion in a continuous time and finite velocity strong memory model \cite{han2021self}, even in the presence of rests \cite{fedotov2022superdiffusion}. However, when including a trapping state, the superdiffusion caused by reinforcement was only transient \cite{han2021anomalous}. 
		
		In biology, cell motility and intracellular transport often exhibit anomalous characteristics and memory effects \cite{bressloff2013stochastic,ariel2015swarming,fedotov2018memory,han2020deciphering}.
	The movement of organelles is often subdiffusive due to the crowded cytoplasm \cite{,han2020deciphering}, which is in direct contrast with the need to efficiently and quickly transport material to specific targets, accomplished by active transport. 		
	Apart from this,  cellular populations  are almost always heterogeneous \cite{altschuler2010cellular}, an example being the different molecular expression levels across individual cells in the brain \cite{muotri2005somatic,lomvardas2006interchromosomal,coufal2009l1}. Furthermore, single cells contain $\sim10^2-10^3$ heterogeneous vesicles with various sizes, morphology and motion essential for all eukaryotic life such as lysosomes \cite{zhu2021metabolomic}. 
		Mathematically, models accounting for static population heterogeneity need to be explored so that the `population-averaged assays' \cite{altschuler2010cellular}, which pervade much biological literature \cite{han2020deciphering,ba2018whole}, can be accurately quantified and the effects of small yet important subpopulations properly identified \cite{altschuler2010cellular}.


 The aim of this paper is to explore the effects of population heterogeneity, characterized by a distribution in transition probability, on the fractional master equation. Below, we demonstrate how heterogeneity changes the fundamental characteristic of the fractional master equation, used in modelling many biological processes that exhibit anomalous trapping \cite{fedotov2021variable,fox2021aging}. The effective underlying random walk exhibits self-reinforcement due to the ensemble averaged conditional transition rates increasing as previous steps accumulate. Moreover, by introducing heterogeneity into the fractional master equation generates ballistic superdiffusion even when the fractional exponent $\mu \rightarrow 1$. We show from a random walk perspective the reason behind why heterogeneity is needed in natural phenomena for efficient transport of an ensemble. Furthermore, we show the mathematical link between population heterogeneity and strong memory. 
 While the topic of random walks in heterogeneous, random environments have been covered extensively in literature \cite{hughes1995random,hughes1996random}, it will not be treated in this paper.

 \section{Fractional master equation with random transition probabilities}	
	The anomalous movement of particles on a lattice that experience trapping with heavy-tailed waiting times can be described by the fractional master equation \cite{klafter2011first}
	\begin{equation}
		\frac{\partial p}{\partial t} = - i(x,t) + q i(x-a,t) + (1-q) i(x+a,t).
		\label{eq:fractionalmasterequation}
	\end{equation}
	Here $p$ is the probability to find the particle at position $x=ka$ ($k\in\mathbb{Z}$) and time $t$. The anomalous escape rate $i(x,t)$ is defined as
	\begin{equation}
		i(x,t) = \tau_0^{-\mu} \mathcal{D}_t^{1-\mu} p(x,t), \quad 0<\mu<1
		\label{eq:integralescape}
	\end{equation}
	and $\mathcal{D}_t^{1-\mu}$ is the Riemann-Liouville derivative
	\begin{equation}
		\mathcal{D}_t^{1-\mu} p(x,t)=\frac{1}{\Gamma(\mu)}\frac{\partial}{\partial t}\int_{0}^{t}\frac{p(x,t')}{(t-t')^{1-\mu}}dt'.
		\label{eq:riemannliouvillederivative}
	\end{equation}
	 Equation \eqref{eq:fractionalmasterequation} describes a random walk where a particle leaves its current state $x$ at time $t$ with rate $i(x,t)$ and either jumps with constant probability $q$ or $1-q$ to $x+a$ or $x-a$, respectively \cite{klafter2011first}. 
	The anomalous rate defined in Eq. \eqref{eq:integralescape} characterizes waiting times that are Mittag-Leffler distributed \cite{fedotov2019asymptotic}. From \eqref{eq:fractionalmasterequation}, by setting $q = 1/2$ and taking the continuous space limit, one can obtain the fractional diffusion equation $\partial p/\partial t = D_{\mu}\partial^2 \mathcal{D}_t^{1-\mu}p(x,t)/\partial x^2,$	with the fractional diffusion coefficient $D_\mu = a^2 /2\tau_o^{\mu}$. Equation \eqref{eq:fractionalmasterequation} with $q=1/2$ and the fractional diffusion equation produces subdiffusive behavior characterized by the mean-squared displacement (and also the variance since the mean is zero) $m^{(2)}(t) \sim t^{\mu}$ where $0<\mu<1$. In order to model population heterogeneity, $q$, the probability of jumping one step in the positive direction, now becomes a random variable for each independent realization of a random walk. In this case, what is the behavior of the ensemble average of the heterogeneous population? 
	
	Clearly for many biological processes, such as intracellular transport \cite{bressloff2013stochastic}, the value of $q$ is heterogeneous across the population of particles.
	Since the bias parameter $q$ is related to the speed as $v\sim 2q-1$, therefore, $q$ can be obtained from the speed distribution  in experiments. Population heterogeneity in speeds is evident in many publications on the topic of intracellular transport \cite{flores2011roles,fedotov2018memory,han2020deciphering} and cell motility \cite{ariel2015swarming}. 
	To account for the heterogeneity across a population of particles, consider that $q$ in Eq. \eqref{eq:fractionalmasterequation} is a random variable that is Beta distributed with a probability density function
	\begin{equation}
		f(q) = \frac{q^{\alpha_+-1}(1-q)^{\alpha_- -1}}{B(\alpha_+,\alpha_-)}
		\label{eq:qBetaPDF}
	\end{equation}
	where $B(\alpha_+,\alpha_-)$ is the beta function.
	
	If $q$ becomes random, how does the anomalous behavior in Eq. \eqref{eq:fractionalmasterequation} change? One might reasonably expect that ensemble fluctuations in $q$ will increase the dispersion of particles leading to randomness of the fractional diffusion coefficient. This idea for standard diffusion has been considered by theories of `diffusing diffusivity' \cite{wang2012brownian,chubynsky2014diffusing,jain2016diffusing} and such heterogeneity was demonstrated to be advantageous for biochemical processes triggered by first arrival \cite{lanoiselee2018diffusion}. Moreover, heterogeneity can be modelled in many ways such as a non-constant diffusion coefficient \cite{metzler2020superstatistics,grebenkov2021exact,sandev2022heterogeneous} or a non-constant anomalous exponent \cite{chechkin2005fractional,korabel2010paradoxes,berry2014spatial,fedotov2019asymptotic,fedotov2021variable}. Dichotomously alternating force fields in the fractional Fokker-Planck equation have also been used to model temporal heterogeneity  \cite{heinsalu2007use}.
	
	In what follows, we will demonstrate that the randomness of $q$ leads to the phenomenon of \textit{ensemble self-reinforcement} and is also connected to random walks exhibiting strong memory. To show this, we need to find the explicit expression for the ensemble averaged probability $\bar{p}(x,t)$ in continuous time 	defined as
	\begin{equation}
		\bar{p}(x,t) = \int_{0}^{1} p(x,t|q) f(q)dq,
		\label{eq:qaverageprobdensity}
	\end{equation}
	where $p(x,t|q)$ is the solution for the master equation \eqref{eq:fractionalmasterequation} with a single value of $q$. In order to do this, we first consider the underlying discrete time random walk for \eqref{eq:fractionalmasterequation} and then utilize the idea of subordination \cite{feller1971introduction, klafter2011first}.
	
	
	
	
	
	
	
	

	\section{Ensemble self-reinforcement and strong memory effects} The underlying discrete time random walk for Eq. \eqref{eq:fractionalmasterequation} is described by the difference equation 
	\begin{equation}
		X_{n+1}=X_n+\xi_{n+1}
	\end{equation}
	where the random jump $\xi_n = \pm a$ with probability $q$ and $1-q$ respectively, and $X_0=0$.  The conditional probability \begin{equation}
		P(x,n|q)=\text{Prob}\{ X_n=x \}
	\end{equation}
	obeys the master equation
	\begin{equation}
		P(x,n+1|q) = qP(x-a,n|q) +(1-q)P(x+a,n|q).
		\label{eq:discretedistributionmaster1}
	\end{equation}
	The solution \cite{feller1971introduction} is
	\begin{equation}
		P(x,n|q)= \binom{n}{\frac{1}{2}(n+\frac{x}{a})}
		q^{\frac{1}{2}(n+\frac{x}{a}) }(1-q)^{\frac{1}{2}(n-\frac{x}{a}) } .
		\label{eq:discretesolution}
	\end{equation}
	The particle reaches the point $x$ at time $n$ if it makes $\frac{1}{2}(n+x/a)$ positive jumps and $\frac{1}{2}(n-x/a)$ negative jumps.
	
	
	Next, we define a probability function
	\begin{equation}
		\bar{P}(x,n) = \int_{0}^{1} P(x,n|q) f(q) dq,
		\label{eq:qaveragedprobability}
	\end{equation}
	which describes the effective underlying random walk for $\bar{X}_n$ such that
	\begin{equation}
		\bar{P}(x,n) = \text{Prob}\{\bar{X}_n=x\}.
	\end{equation}
	By averaging \eqref{eq:discretedistributionmaster1} using $f(q)$ from \eqref{eq:qBetaPDF}, we obtain the master equation 
	\begin{equation}
		\bar{P}(x,n+1)=u_{n}^{+}(x-a) \bar{P}(x-a,n) +u_{n}^{-}(x+a)\bar{P}(x+a,n) 
		\label{eq:discretemaster}
	\end{equation}
	where the transition probabilities $u_{n}^{+}(x)$ and $u_{n}^{-}(x)$ are defined as follows
	\begin{equation}
		u_{n}^{+}(x)=\frac{\int_{0}^{1} q  P(x,n|q) f(q)dq }{\int_{0}^{1} P(x,n|q) f(q)dq }, \quad
		u_{n}^{-}(x)=1-u_{n}^{+}(x).
		\label{eq:utransition}
	\end{equation}
	Transition probabilities \eqref{eq:utransition} follow from averaging \eqref{eq:discretedistributionmaster1} with respect to $f(q)$. By using the solution \eqref{eq:discretesolution} we find
	\begin{equation}
		u_{n}^{\pm}(x)=\frac{\alpha_{\pm} +\frac{1}{2}\left(n\pm\frac{x}{a}\right)}{\alpha_++\alpha_- +n}.
		\label{eq:utransition2}
	\end{equation}

    Surprisingly, randomness of the parameter $q$ generates effective transition probabilities, $u_n^{\pm}(x)$, which describes the \textit{ensemble self-reinforcement} phenomenon. It follows from \eqref{eq:utransition2} that the probability to step in the positive or negative direction increases as more steps in those directions are made in the past, which is known as self-reinforcement.
    In what follows, we demonstrate the link between Eqs. \eqref{eq:discretemaster} with \eqref{eq:utransition2} and random walks with transition probabilities dependent on the entire history of its past, a property called strong memory. Furthermore, we provide an explanation to how these two concepts are linked despite the difference in the underlying mechanism.
	
	In fact, Eq. \eqref{eq:discretemaster} describes a random walk with strong memory: $\bar{X}_{n+1} = \bar{X}_n + \bar{\xi}_{n+1}$. 
 	The conditional transition probability for the discrete steps, $\bar{\xi}_n$, depends on its entire history such that
	\begin{equation}
	   \text{Prob}\{\bar{\xi}_{n+1}=\pm a\hspace{0.1cm}|_{\bar{\xi}_1,\cdots,\bar{\xi}_n}\} = \frac{\alpha_{\pm}+n_{\pm}}{\alpha_++\alpha_-+n}.
	   \label{eq:xidefinition_strongmemory_asymmetric}
	\end{equation}
	Here $n_{\pm}$ is the number of steps taken in the positive and negative directions, respectively. Equation \eqref{eq:xidefinition_strongmemory_asymmetric} can be obtained from the transition probabilities \eqref{eq:utransition2} by combining the current position $x=a(n_+-n_-)$ and the total number of steps $n=n_++n_-$. The transition probabilities \eqref{eq:xidefinition_strongmemory_asymmetric} depend on the entire history because $n_{\pm}$ counts the number of steps taken in the positive and negative directions up to time $n$. This dependence of the conditional transition probability on the entire history of the random walk is known in the literature as strong memory \cite{schutz2004elephants,kumar2010memory,da2013non,budini2016inhomogeneous,baur2016elephant,pemantle1988phase,han2021self,fedotov2022superdiffusion}. The conditional transition probability \eqref{eq:xidefinition_strongmemory_asymmetric} is exactly the same as that of a P\'olya urn model \cite{feller1971introduction,pemantle1988phase} where initially the urn contains $\alpha_+$ red and $\alpha_-$ black balls and then only one ball is added per draw with $n_{\pm}$ the number of red and black balls drawn, respectively. 
	
	
	Comparing Eqs. \eqref{eq:utransition2} and \eqref{eq:xidefinition_strongmemory_asymmetric}, it is clear that \textit{ensemble self-reinforcement} generates strong memory effects. However, a key feature of the random walk governed by \eqref{eq:discretemaster} is that the strong memory effect is a by-product of the heterogeneity in the ensemble. Does this mean that, through heterogeneity, particles performing the random walk in \eqref{eq:discretemaster} are somehow more likely to step in the positive or negative direction dependent on their history? On the contrary, this \textit{ensemble self-reinforcement} is a consequence of sampling a heterogeneous population. This type of effect that leads to reinforcement is discussed in probability theory as aftereffect or spurious contagion \cite{feller1971introduction}. Rather than steps becoming more likely given the previous step, particles with a very high propensity to always step to the right or left are more likely to be found at the positive or negative extremities of the population. This is especially pertinent in cell biology as often in microscopic scales, such as intracellular organelles, there is no internal mechanism of reinforcement or `contagion' and memory effects could be due to sampling a heterogeneous population. 	Equations \eqref{eq:utransition2}  illustrate the fact that simply changing the transition probability $q$ from a constant to a random variable completely changes the fundamental underlying mechanism of transitions in the ensemble. 
	
	
	
	\section{Ensemble averaged solution for the fractional master equation}
	By using the concept of subordination \cite{feller1971introduction,klafter2011first}, we can find the explicit expression for the ensemble averaged probability distribution $\bar{p}(x,t)$ in continuous time defined in \eqref{eq:qaverageprobdensity}. The underlying random walk for the master equation \eqref{eq:fractionalmasterequation} is the compound fractional Poisson process \cite{laskin2003fractional}
	\begin{equation}
	    X_{\mu}(t) =  \sum_{i=1}^{N_{\mu}(t)}\xi_i
	\end{equation}
	where $\xi_i$ are random jumps, $N_\mu(t)$ is the fractional Poisson process and $X_{\mu}(0) = 0.$ The latter describes the number of steps taken at time $t$ given the waiting time is Mittag-Leffler distributed \cite{laskin2003fractional}.
	Using subordination \cite{feller1971introduction,klafter2011first}, we can write
	\begin{equation}
		\bar{p}(x,t)=\sum_{n=0}^{\infty}\bar{P}(x,n)Q_{\mu}(n,t)
		\label{eq:steps_time_convolution_distribution}
	\end{equation}
	where $\bar{P}(x,n)$ is defined in \eqref{eq:qaveragedprobability} and $Q_{\mu}(n,t)= \text{Prob}\{N_\mu(t)=n\}$. One can also write down $\bar{p}(x,t)$ in terms of the position of the continuous-time random walk
	\begin{equation}
		\bar{p}(x,t) = \text{Prob}\{\bar{X}_{\mu}(t)=x\} 
	\end{equation}
	where 
	\begin{equation}
		\bar{X}_{\mu}(t) = \bar{X}_{N_{\mu}(t)}.
		\label{eq:xbarmu definition}
	\end{equation}
	From the master equation \eqref{eq:discretemaster} or by averaging the solution \eqref{eq:discretesolution} as shown in \eqref{eq:qaveragedprobability}, one can obtain
	\begin{equation}
		\bar{P} = \binom{n}{\frac{1}{2}(n+\frac{x}{a})}\frac{B\left(\frac{1}{2}(n+\frac{x}{a})+\alpha_+ , \frac{1}{2}(n-\frac{x}{a})+\alpha_- \right) }{B(\alpha_-,\alpha_+)}.
		\label{eq:discretesolution2}
	\end{equation}
	The probability $Q_{\mu}(n,t)$ is given by \cite{laskin2003fractional},
	\begin{equation}
		Q_{\mu}(n,t)=\left(\frac{t}{\tau_0}\right)^{n\mu} \sum_{k=0}^{\infty}\frac{(k+n)!}{n!k!}\frac{\left(-\frac{t}{\tau_0}\right)^{k\mu}}{\Gamma(\mu(k+n)+1)}.
		\label{eq:fractionalPoissonSolution}
	\end{equation}
	So substituting \eqref{eq:discretesolution2} and \eqref{eq:fractionalPoissonSolution} into \eqref{eq:steps_time_convolution_distribution} gives the ensemble averaged solution of the fractional master equation \eqref{eq:fractionalmasterequation} through subordination involving the fractional Poisson process and the underlying discrete random walk with self-reinforcement.

\begin{figure}
		\centering
		\includegraphics[width=\linewidth]{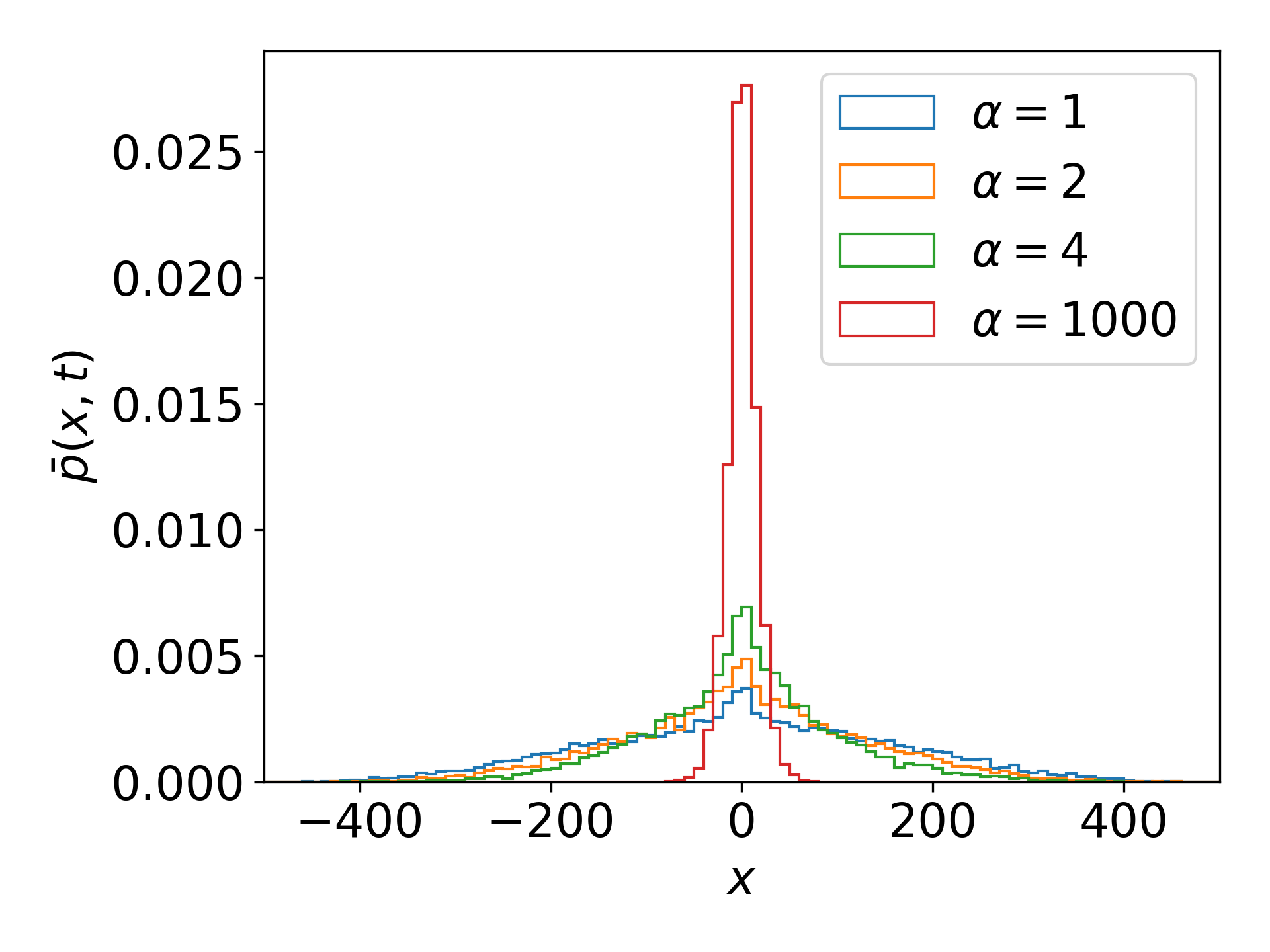}
		\caption{Probability distribution of random walkers in continuous time with Mittag-Leffler distributed waiting times $\mu=0.75$, $\tau_0=1$, varying values of $\alpha_+=\alpha_-=\alpha/2$ for the Beta distribution \eqref{eq:qBetaPDF}, $a=1$, $t_{end}=10^3$ and $N=10^4$.}
		\label{fig:pdfml}
	\end{figure}

	Fig. \ref{fig:pdfml} illustrates the solution \eqref{eq:steps_time_convolution_distribution} obtained by Monte Carlo simulations for the symmetrical case ($\alpha_+=\alpha_-$).
	One can see the unusually strong dispersion for the subdiffusive master equation, which is a result of the interaction between \textit{ensemble self-reinforcement}  described by $\bar{P}(x,n)$ and heavy-tailed waiting times with divergent mean described by $Q_{\mu}(n,t)$.
	
	\subsection{Monte Carlo simulations} The simulations for all figures were performed in the following way:
	\begin{enumerate}
		\item Initialize $N$ particles at $X(0) = 0$. For each particle, the value of $q$ is a random variable drawn from a Beta distribution.
		\item Then, for each particle, draw a value for $T$ from Mittag-Leffler distributed random numbers. Then $X(t+T) = X(t) +Z$ where $\text{Prob}[Z = 1] = q$ and $\text{Prob}[Z = -1] = 1-q$.
		\item Iterate until a required end time $t_{end}$.
	\end{enumerate}
	Mittag-Leffler distributed random numbers were generated using the standard procedure (see (20) in \cite{fulger2008monte} or \cite{kozubowski1999univariate}).

	\section{Superdiffusion generated by ensemble self-reinforcement} Now, we will show how ballistic superdiffusion can arise due to \textit{ensemble self-reinforcement}. To do this, we need to find the moments corresponding to the discrete case of \eqref{eq:steps_time_convolution_distribution} 
	\begin{equation}
		M^{(m)}(n)= \sum_{x\in\Omega} x^m \bar{P}(x,n), \quad m\in\{1,2,\cdots\}.
		\label{eq:momentsgeneral}
	\end{equation}
	Here, the summation is over all the lattice positions $\Omega = \{ka\}$ with $k\in\mathbb{Z}$.
	Using \eqref{eq:qaveragedprobability}, we can rewrite \eqref{eq:momentsgeneral} as
	\begin{equation}
		M^{(m)}(n) = \int_{0}^{1} \left[ \sum_{x\in\Omega} x^m P(x,n|q) \right] f(q) dq.
		\label{eq:momentsrearragned}
	\end{equation}
	Recognizing that the summation in \eqref{eq:momentsrearragned} is simply the m\textsuperscript{th} moment of the discrete random walk, $X_n$, governed by \eqref{eq:discretedistributionmaster1} for any fixed value of $q$, we find
	\begin{equation}
		M^{(m)}(n) = \int_0^1\mathbb{E}[(X_n)^m]f(q)dq.
		\label{eq:momentsexpectation}
	\end{equation}
	First, we find the conditional moments of the underlying random walk for fixed $q$: $\mathbb{E}(X_n)=G'(1)$ and $\mathbb{E}(X_n^2) =G''(1)+G'(1)$, where $G(z) = \left[qz^a+(1-q)z^{-a}\right]^n$ is the probability generating function \cite{coxmillertextbook}. Performing this calculation, we obtain
	\begin{equation}
		\mathbb{E}(X_n) = an(2q-1)
		\label{eq:firstmomentdiscrete}
	\end{equation}
	and 
	\begin{equation}
		\mathbb{E}(X_n^2) = a^2(2q-1)^2n^2 +(1-(2q-1)^2)a^2n.
		\label{eq:secondmomentdiscrete}
	\end{equation}
	The variance is proportional to $n$:
	\begin{equation}
		\text{Var}[X_n] = (1-(2q-1)^2)a^2 n .
		\label{eq:variancediscrete}
	\end{equation}
	Now, we take the average  of \eqref{eq:firstmomentdiscrete} and \eqref{eq:secondmomentdiscrete} to obtain the variance of the effective random walk.
	In contrast to \eqref{eq:variancediscrete}, the variance involves a term proportional to $n^2$
	\begin{equation}
		\begin{split}
			\text{Var}\left[\bar{X}_n \right] =& (\overline{(2q-1)^2}-(2\overline{q}-1)^2)a^2 n^2 \\
			& +(1-\overline{(2q-1)^2})a^2n
		\end{split}
	\label{eq:variance_xbar_n}
	\end{equation}
	where 
	\begin{equation}
			\bar{q} =\int_{0}^{1}qf(q)dq,
			\hspace{0.4cm}
			\overline{(2q-1)^2}	= \int_{0}^{1}(2q-1)^2f(q)dq.
	 \end{equation}
	The difference between \eqref{eq:variancediscrete} and \eqref{eq:variance_xbar_n} is fundamentally important because the term proportional to $n^2$ generates ballistic superdiffusion
	
	\subsection{Symmetric Beta distribution: zero average advection}


	To avoid the averaged advection caused by an asymmetric Beta distribution, we only consider cases when the Beta distribution is symmetric, 
	\begin{equation}
		\alpha_+ = \alpha_- = \frac{\alpha}{2}.
	\end{equation}
	The absence of averaged advection is emphasised in Fig. \ref{fig:pdfml}, which shows symmetric distributions for different values of $\alpha$. Figure \ref{fig:pdfml} also shows that in the limit of large $\alpha$, the distribution reverts back to the distribution typical for the subdiffusive regime. 
		
	For the symmetric case with $\overline{q} = 1/2$, one can obtain
	\begin{equation}
		\text{Var}[\bar{X}_n] = M^{(2)}-\left[M^{(1)}\right]^2=  \frac{a^2}{1+\alpha}n^2 + \frac{a^2\alpha}{1+\alpha}n.
		\label{eq:qaveragedmoments}
	\end{equation}

	The reason why the variance has a term proportional to $n^2$ can be explained by \textit{ensemble self-reinforcement} expressed by the transition probabilities in \eqref{eq:utransition2}, which leads to a greater dispersion of particles over time compared to standard random walks. Note that this result can be obtained by also finding the moments through a recursion relation from the master equation \eqref{eq:discretemaster} \cite{budini2016inhomogeneous}.
	
\begin{figure}
		\centering
		\includegraphics[width=\linewidth]{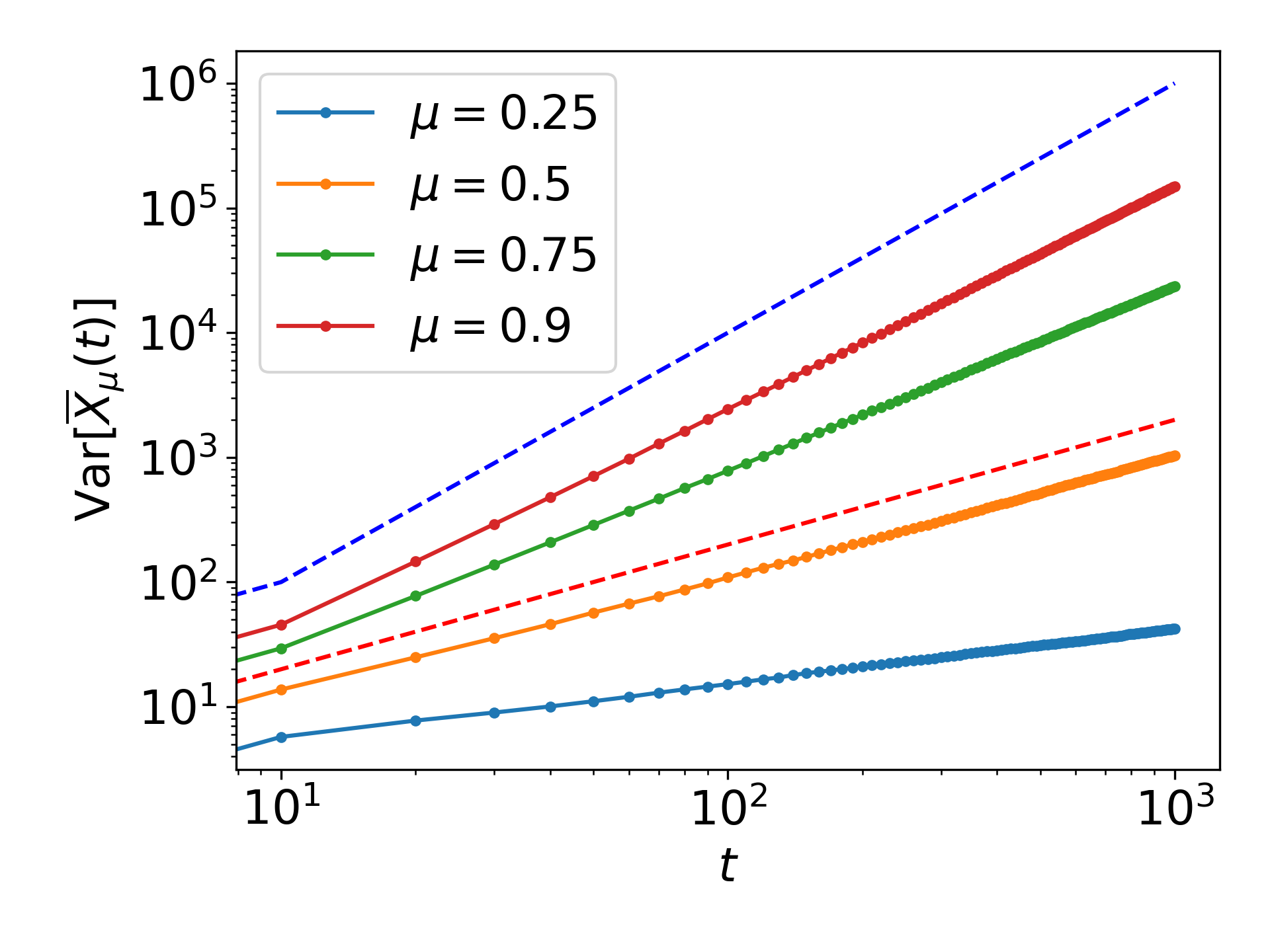}
		\caption{Variance of random walkers in continuous time with Mittag-Leffler distributed waiting times with varying values of $\mu$, $\tau_0=1$, $\alpha/2=1/2$, $t_{end}=10^3$ and $N=10^4$. Blue dashed line shows $\text{Var}\left[ \bar{X}_{\mu}(t) \right]\propto t^2$. Red dashed line shows $\text{Var}\left[ \bar{X}_{\mu}(t) \right]\propto t$.}
		\label{fig:msdml_varyingmu}
	\end{figure}

\begin{figure}
		\centering
		\includegraphics[width=0.95\linewidth]{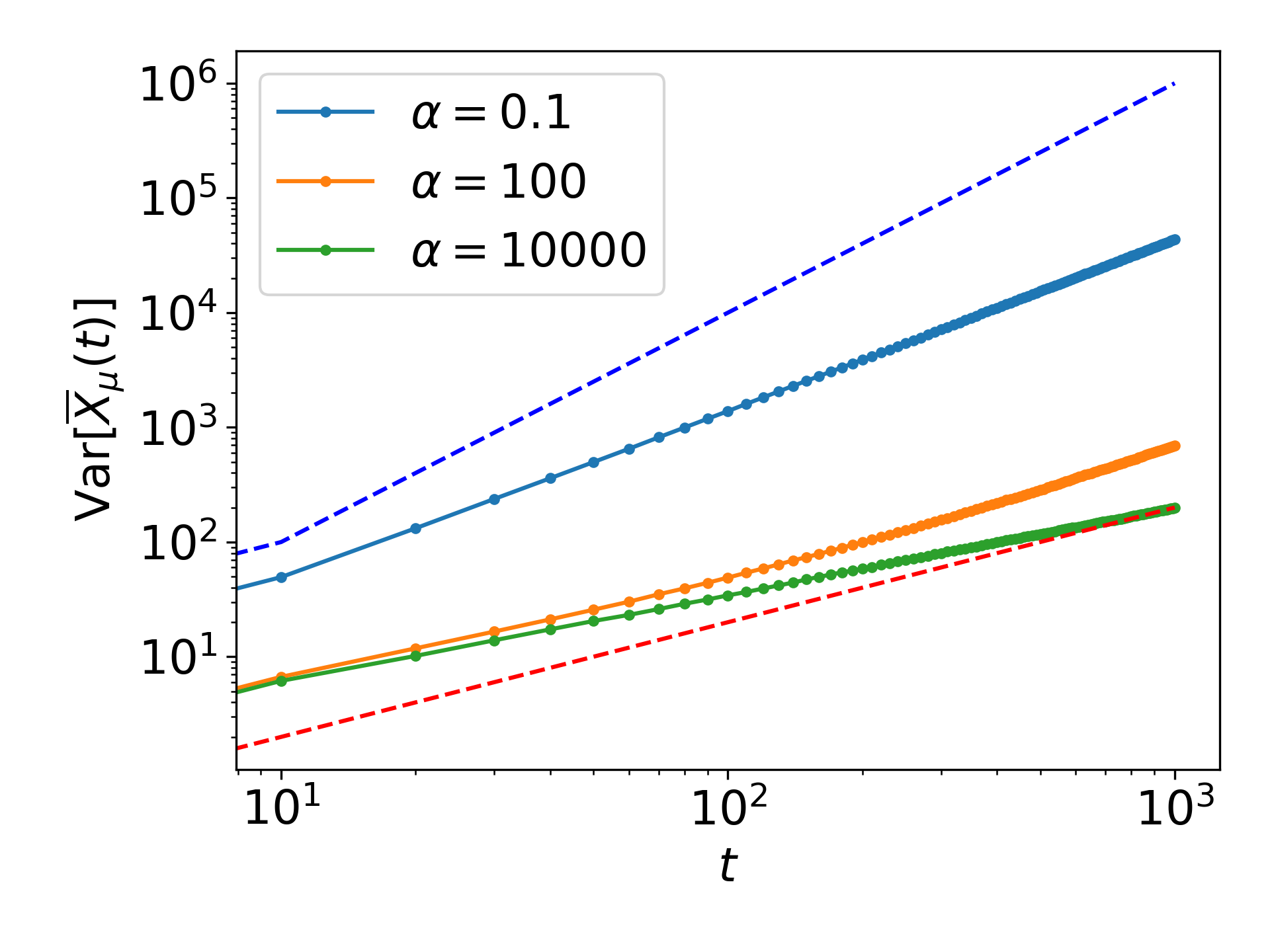}
		\caption{Variance of random walkers in continuous time with varying values of $\alpha/2$, Mittag-Leffler distributed waiting times with $\mu=0.75$, $\tau_0=1$, $t_{end}=10^3$ and $N=10^4$. Blue dashed line shows $\text{Var}\left[ \bar{X}_{\mu}(t) \right]\propto t^2$. Red dashed line shows $\text{Var}\left[ \bar{X}_{\mu}(t) \right]\propto t$.}
		\label{fig:msdml_varyingalpha}
	\end{figure}

	One can find the variance for the effective continuous-time random walk \begin{equation}
		\text{Var}\left[\bar{X}_{\mu}(t)\right] = \frac{a^2}{1+\alpha}\langle n^2(t)\rangle + \frac{a^2 \alpha}{1+\alpha} \langle n(t)\rangle
		\label{eq:altsecondmomentdiscrete}
	\end{equation}
	where $\bar{X}_{\mu}(t)$ is defined in \eqref{eq:xbarmu definition}, $\langle n^2(t)\rangle$ and $\langle n(t)\rangle$ are derived from the fractional Poisson process \cite{laskin2003fractional} as 
	\begin{equation}
		\langle n(t)\rangle = \frac{1}{\Gamma(\mu+1)}\left(\frac{t}{\tau_0}\right)^{\mu}
	\end{equation}
	and
	\begin{equation}
		\langle n^2(t)\rangle = \frac{1}{\Gamma(\mu+1)}\left(\frac{t}{\tau_0}\right)^{\mu} + \frac{A_{\mu}}{\Gamma(\mu+1)}\left(\frac{t}{\tau_0}\right)^{2\mu},
	\end{equation}
	where
	\begin{equation}
		A_{\mu}=\frac{\sqrt{\pi}}{2^{2\mu-1}\Gamma(\mu+\frac{1}{2})} = \frac{\Gamma(\mu)}{\Gamma(2\mu)}.
	\end{equation}
	Finally, the variance in continuous time is
	\begin{equation}
		\text{Var}[\bar{X}_{\mu}(t)]= \frac{a^2}{\Gamma(\mu+1)}\left[  \left(\frac{t}{\tau_0}\right)^{\mu}+\frac{A_{\mu}}{1+\alpha}\left(\frac{t}{\tau_0}\right)^{2\mu}\right].
		\label{eq:secondmomentcontinuous}
	\end{equation}
	The appearance of superdiffusion is demonstrated by numerical simulations in Figs. \ref{fig:msdml_varyingmu} and \ref{fig:msdml_varyingalpha}. Figure \ref{fig:msdml_varyingmu} demonstrates numerically the relation in \eqref{eq:secondmomentcontinuous} and \eqref{eq:variancePoisson} since for values of $\mu<0.5$, $\text{Var}\left[ \bar{X}_{\mu}(t) \right]$ shows subdiffusion and for values $\mu>0.5$ shows superdiffusion. Moreover for $\mu=0.5$, $\text{Var}\left[\bar{X}_{\mu}(t)\right]$ is exactly diffusive.
	Note that when $\mu=1$, $N_{\mu}(t)$ becomes a Poisson process with rate $1/\tau_0$ and the variance becomes ballistic:
	\begin{equation}
		\text{Var}[\bar{X}_{\mu}(t)] =  a^2\frac{t}{\tau_0} + \frac{a^2}{1+\alpha}\left(\frac{t}{\tau_0}\right)^{2}
		\label{eq:variancePoisson}
	\end{equation}
	This result is different from the case when an external force combines with the fractional master equation \cite{metzler1998anomalous,goychuk2006current} where the first moment is $m^{(1)}(t)\sim t^{\mu}$ and so the second moment becomes $m^{(2)}(t)\sim t^{2\mu}$. The superdiffusion caused in this new process is a result of a heterogeneous population of particles and this generates \textit{ensemble self-reinforcement} demonstrated by \eqref{eq:utransition2}.
		A simple random walk with bias and fractional rates would be described by \eqref{eq:fractionalmasterequation} where $q$ is a constant. 
		Explicitly, the mean position and variance of this random walk conditional on the transition probability are \cite{shlesinger1974asymptotic,metzler1998anomalous,goychuk2006current}
		\begin{equation}
			\mathbb{E}[X(t)|q] = \frac{a(2q-1)}{\Gamma(\mu+1)}\left( \frac{t}{\tau_0} \right)^{\mu},
			\label{eq:mean_averagelate}
		\end{equation}
		\begin{equation}
			\begin{split}
				\text{Var} \left[ X(t) | q\right] =& (2q-1)^2\left( \frac{t}{\tau_0} \right)^{2\mu}\left[ \frac{2a^2}{\Gamma(2\mu+1)} - \frac{a^2}{\Gamma(\mu+1)^2} \right]\\ &+ \frac{a^2}{\Gamma(\mu+1)}\left( \frac{t}{\tau_0} \right)^{\mu}.
			\end{split}
			\label{eq:variance_averagelate}
		\end{equation}
		
		Clearly, \eqref{eq:variance_averagelate} exhibits superdiffusive behavior but the terms proportional to $t^{2\mu}$ disappear when $\mu=1$. The reason for this is that the underlying random walk model, $X_n$, has variance proportional to $n$, as seen in \eqref{eq:variancediscrete}. However, \eqref{eq:secondmomentcontinuous} exhibits ballistic superdiffusion when $\mu=1$ because the effective random walk of the ensemble, $\bar{X}_n$, has variance \eqref{eq:secondmomentdiscrete} proportional to $n^2$ and $n$.
		
	
	
	
	
	Furthermore, from this new, heterogeneous population model we are able to achieve a smooth transition in time between subdiffusion and superdiffusion. This is evident by increasing the value of $\alpha\rightarrow\infty$. This is intuitive as the symmetric Beta distribution approaches a delta function centered at $q=1/2$ as $\alpha\rightarrow\infty$ and so we recover the standard fractional master equation and the resulting subdiffusion. However when $\alpha \sim 1$ and $\mu>1/2$, we obtain superdiffusion in the long-time limit. This transition between superdiffusion and subdiffusion is demonstrated using computational simulations in Fig. \ref{fig:msdml_varyingalpha}.
	
	\section{Discussion}
	Although there is vast literature on strong memory effects in statistical physics \cite{schutz2004elephants,kumar2010memory,da2013non,budini2016inhomogeneous,baur2016elephant,pemantle1988phase,han2021self,fedotov2022superdiffusion}, many elephant random walk-like models lack the mechanism of how the strong memory is produced.
	Given that a heterogeneous population of random walkers emulates strong memory, this opens a new avenue for modelling biological processes that display strong memory properties and yet are heterogeneous ensembles of inanimate objects, like organelles and micromolecules. Might it be that nature has developed a mechanism like \textit{ensemble self-reinforcement} that we demonstrate in \eqref{eq:utransition2} as a proxy for strong memory? Such questions have plagued the field of intracellular transport for decades where brainless membrane-bound vesicles seemingly engage in random walks that appear to have correlations caused by strong memory effects \cite{chen2015memoryless,fedotov2018memory}. For example, a high value of $q$ might represent a higher affinity to attach to the dynein family of motor proteins and therefore the particle moves very directionally towards the cell nucleus whereas a low value of $q$ would be a higher affinity to attach to kinesin which moves towards the cell periphery. A value of $q \sim 1/2$ would imply that a particle may have equal chance to move in either direction.s
		The relationship between $q$ and speed $v$ of a vesicle is $v\sim 2q-1$. So the bias parameter $q$ can be obtained from experiments.
		Heterogeneity in velocities of intracellular vesicles is well-established \cite{flores2011roles,fedotov2018memory,han2020deciphering}.
	\textit{Ensemble self-reinforcement} enables the organization of directional movement as an ensemble effect from heterogeneity. Furthermore, we showed that \textit{ensemble self-reinforcement} can generate ballistic superdiffusion.
	
	This finding also fits nicely with the emerging theory that, in biological processes, the first arrival times of a signal to a cell (or neuron) influence the subsequent system behavior far more than the average arrival times \cite{schuss2019redundancy}. With \textit{ensemble self-reinforcement} the cell can organize the movement of these particles such that it maintains efficiency of transport and overcomes the trapping that occurs in the crowded cytoplasm. We hypothesise that \textit{ensemble self-reinforcement} is a way that the cell efficiently transports vesicles in a heavily crowded intracellular environment, which has been shown to be subdiffusive \cite{ba2018whole,fedotov2021variable}.
	
	\section{Summary} 
In this paper, we formulate a fractional master equation with random transition probabilities across the populations of random walkers. This population heterogeneity generates ensemble average transition probabilities that that increase with the number of steps taken previously, which we call \textit{ensemble self-reinforcement}. These averaged transition probabilities open a new avenue to model strong memory effects through a heterogeneous ensemble of random walkers. Furthermore, we show analytical solutions for the variance and probability density function of the ensemble averaged, effective random walk.

Through this, we establish the connection between random walks with a heterogeneous ensemble and those with strong memory where the transition probability depends on the entire history of steps. 
		We find the ensemble averaged solution of the fractional master equation through subordination involving the fractional Poisson process counting the number of steps at a given time and the underlying discrete random walk with self-reinforcement. 
		We also find the exact solution for the variance which exhibits superdiffusion even as the fractional exponent tends to 1.
	This paper demonstrates that heterogeneous populations of anomalous random walks can achieve effective transition probabilities describing strong memory, which we call \textit{ensemble self-reinforcement}. We find that such heterogeneous populations overcomes heavy-tailed waiting times with divergent mean to exhibit ensemble superdiffusion thus revealing an intrinsic advantage of heterogeneity. Moreover, this provides a new mechanism through which seemingly unintelligent systems can exhibit strong memory. 
	
	\bibliography{real}

\begin{thebibliography}{65}
\expandafter\ifx\csname natexlab\endcsname\relax\def\natexlab#1{#1}\fi
\expandafter\ifx\csname bibnamefont\endcsname\relax
  \def\bibnamefont#1{#1}\fi
\expandafter\ifx\csname bibfnamefont\endcsname\relax
  \def\bibfnamefont#1{#1}\fi
\expandafter\ifx\csname citenamefont\endcsname\relax
  \def\citenamefont#1{#1}\fi
\expandafter\ifx\csname url\endcsname\relax
  \def\url#1{\texttt{#1}}\fi
\expandafter\ifx\csname urlprefix\endcsname\relax\def\urlprefix{URL }\fi
\providecommand{\bibinfo}[2]{#2}
\providecommand{\eprint}[2][]{\url{#2}}

\bibitem[{\citenamefont{Bouchaud and Georges}(1990)}]{bouchaud1990anomalous}
\bibinfo{author}{\bibfnamefont{J.-P.} \bibnamefont{Bouchaud}} \bibnamefont{and}
  \bibinfo{author}{\bibfnamefont{A.}~\bibnamefont{Georges}},
  \bibinfo{journal}{Physics reports} \textbf{\bibinfo{volume}{195}},
  \bibinfo{pages}{127} (\bibinfo{year}{1990}).

\bibitem[{\citenamefont{Metzler et~al.}(1999)\citenamefont{Metzler, Barkai, and
  Klafter}}]{metzler1999anomalous}
\bibinfo{author}{\bibfnamefont{R.}~\bibnamefont{Metzler}},
  \bibinfo{author}{\bibfnamefont{E.}~\bibnamefont{Barkai}}, \bibnamefont{and}
  \bibinfo{author}{\bibfnamefont{J.}~\bibnamefont{Klafter}},
  \bibinfo{journal}{Physical review letters} \textbf{\bibinfo{volume}{82}},
  \bibinfo{pages}{3563} (\bibinfo{year}{1999}).

\bibitem[{\citenamefont{Metzler and Klafter}(2004)}]{metzler2004restaurant}
\bibinfo{author}{\bibfnamefont{R.}~\bibnamefont{Metzler}} \bibnamefont{and}
  \bibinfo{author}{\bibfnamefont{J.}~\bibnamefont{Klafter}},
  \bibinfo{journal}{Journal of Physics A: Mathematical and General}
  \textbf{\bibinfo{volume}{37}}, \bibinfo{pages}{R161} (\bibinfo{year}{2004}).

\bibitem[{\citenamefont{Scalas}(2006)}]{scalas2006application}
\bibinfo{author}{\bibfnamefont{E.}~\bibnamefont{Scalas}},
  \bibinfo{journal}{Physica A: Statistical Mechanics and its Applications}
  \textbf{\bibinfo{volume}{362}}, \bibinfo{pages}{225} (\bibinfo{year}{2006}).

\bibitem[{\citenamefont{Henry et~al.}(2010)\citenamefont{Henry, Langlands, and
  Straka}}]{henry2010fractional}
\bibinfo{author}{\bibfnamefont{B.~I.} \bibnamefont{Henry}},
  \bibinfo{author}{\bibfnamefont{T.}~\bibnamefont{Langlands}},
  \bibnamefont{and} \bibinfo{author}{\bibfnamefont{P.}~\bibnamefont{Straka}},
  \bibinfo{journal}{Physical review letters} \textbf{\bibinfo{volume}{105}},
  \bibinfo{pages}{170602} (\bibinfo{year}{2010}).

\bibitem[{\citenamefont{Klafter and Sokolov}(2011)}]{klafter2011first}
\bibinfo{author}{\bibfnamefont{J.}~\bibnamefont{Klafter}} \bibnamefont{and}
  \bibinfo{author}{\bibfnamefont{I.~M.} \bibnamefont{Sokolov}},
  \emph{\bibinfo{title}{First steps in random walks: from tools to
  applications}} (\bibinfo{publisher}{OUP Oxford}, \bibinfo{year}{2011}).

\bibitem[{\citenamefont{Mandelbrot and
  Van~Ness}(1968)}]{mandelbrot1968fractional}
\bibinfo{author}{\bibfnamefont{B.~B.} \bibnamefont{Mandelbrot}}
  \bibnamefont{and} \bibinfo{author}{\bibfnamefont{J.~W.}
  \bibnamefont{Van~Ness}}, \bibinfo{journal}{SIAM review}
  \textbf{\bibinfo{volume}{10}}, \bibinfo{pages}{422} (\bibinfo{year}{1968}).

\bibitem[{\citenamefont{Kubo}(1966)}]{kubo1966fluctuation}
\bibinfo{author}{\bibfnamefont{R.}~\bibnamefont{Kubo}},
  \bibinfo{journal}{Reports on progress in physics}
  \textbf{\bibinfo{volume}{29}}, \bibinfo{pages}{255} (\bibinfo{year}{1966}).

\bibitem[{\citenamefont{Zwanzig}(1973)}]{zwanzig1973nonlinear}
\bibinfo{author}{\bibfnamefont{R.}~\bibnamefont{Zwanzig}},
  \bibinfo{journal}{Journal of Statistical Physics}
  \textbf{\bibinfo{volume}{9}}, \bibinfo{pages}{215} (\bibinfo{year}{1973}).

\bibitem[{\citenamefont{Goychuk and H{\"a}nggi}(2007)}]{goychuk2007anomalous}
\bibinfo{author}{\bibfnamefont{I.}~\bibnamefont{Goychuk}} \bibnamefont{and}
  \bibinfo{author}{\bibfnamefont{P.}~\bibnamefont{H{\"a}nggi}},
  \bibinfo{journal}{Physical Review Letters} \textbf{\bibinfo{volume}{99}},
  \bibinfo{pages}{200601} (\bibinfo{year}{2007}).

\bibitem[{\citenamefont{Shlesinger et~al.}(1987)\citenamefont{Shlesinger, West,
  and Klafter}}]{shlesinger1987levy}
\bibinfo{author}{\bibfnamefont{M.~F.} \bibnamefont{Shlesinger}},
  \bibinfo{author}{\bibfnamefont{B.}~\bibnamefont{West}}, \bibnamefont{and}
  \bibinfo{author}{\bibfnamefont{J.}~\bibnamefont{Klafter}},
  \bibinfo{journal}{Physical Review Letters} \textbf{\bibinfo{volume}{58}},
  \bibinfo{pages}{1100} (\bibinfo{year}{1987}).

\bibitem[{\citenamefont{Zaburdaev et~al.}(2015)\citenamefont{Zaburdaev,
  Denisov, and Klafter}}]{zaburdaev2015levy}
\bibinfo{author}{\bibfnamefont{V.}~\bibnamefont{Zaburdaev}},
  \bibinfo{author}{\bibfnamefont{S.}~\bibnamefont{Denisov}}, \bibnamefont{and}
  \bibinfo{author}{\bibfnamefont{J.}~\bibnamefont{Klafter}},
  \bibinfo{journal}{Reviews of Modern Physics} \textbf{\bibinfo{volume}{87}},
  \bibinfo{pages}{483} (\bibinfo{year}{2015}).

\bibitem[{\citenamefont{Livina et~al.}(2005)\citenamefont{Livina, Havlin, and
  Bunde}}]{livina2005memory}
\bibinfo{author}{\bibfnamefont{V.~N.} \bibnamefont{Livina}},
  \bibinfo{author}{\bibfnamefont{S.}~\bibnamefont{Havlin}}, \bibnamefont{and}
  \bibinfo{author}{\bibfnamefont{A.}~\bibnamefont{Bunde}},
  \bibinfo{journal}{Physical review letters} \textbf{\bibinfo{volume}{95}},
  \bibinfo{pages}{208501} (\bibinfo{year}{2005}).

\bibitem[{\citenamefont{Piilo et~al.}(2008)\citenamefont{Piilo, Maniscalco,
  H{\"a}rk{\"o}nen, and Suominen}}]{piilo2008non}
\bibinfo{author}{\bibfnamefont{J.}~\bibnamefont{Piilo}},
  \bibinfo{author}{\bibfnamefont{S.}~\bibnamefont{Maniscalco}},
  \bibinfo{author}{\bibfnamefont{K.}~\bibnamefont{H{\"a}rk{\"o}nen}},
  \bibnamefont{and} \bibinfo{author}{\bibfnamefont{K.-A.}
  \bibnamefont{Suominen}}, \bibinfo{journal}{Physical review letters}
  \textbf{\bibinfo{volume}{100}}, \bibinfo{pages}{180402}
  (\bibinfo{year}{2008}).

\bibitem[{\citenamefont{Wolf et~al.}(2008)\citenamefont{Wolf, Eisert, Cubitt,
  and Cirac}}]{wolf2008assessing}
\bibinfo{author}{\bibfnamefont{M.~M.} \bibnamefont{Wolf}},
  \bibinfo{author}{\bibfnamefont{J.}~\bibnamefont{Eisert}},
  \bibinfo{author}{\bibfnamefont{T.~S.} \bibnamefont{Cubitt}},
  \bibnamefont{and} \bibinfo{author}{\bibfnamefont{J.~I.} \bibnamefont{Cirac}},
  \bibinfo{journal}{Physical review letters} \textbf{\bibinfo{volume}{101}},
  \bibinfo{pages}{150402} (\bibinfo{year}{2008}).

\bibitem[{\citenamefont{Luchnikov et~al.}(2020)\citenamefont{Luchnikov,
  Vintskevich, Grigoriev, and Filippov}}]{luchnikov2020machine}
\bibinfo{author}{\bibfnamefont{I.}~\bibnamefont{Luchnikov}},
  \bibinfo{author}{\bibfnamefont{S.}~\bibnamefont{Vintskevich}},
  \bibinfo{author}{\bibfnamefont{D.}~\bibnamefont{Grigoriev}},
  \bibnamefont{and} \bibinfo{author}{\bibfnamefont{S.}~\bibnamefont{Filippov}},
  \bibinfo{journal}{Physical Review Letters} \textbf{\bibinfo{volume}{124}},
  \bibinfo{pages}{140502} (\bibinfo{year}{2020}).

\bibitem[{\citenamefont{Bressloff and Newby}(2013)}]{bressloff2013stochastic}
\bibinfo{author}{\bibfnamefont{P.~C.} \bibnamefont{Bressloff}}
  \bibnamefont{and} \bibinfo{author}{\bibfnamefont{J.~M.} \bibnamefont{Newby}},
  \bibinfo{journal}{Reviews of Modern Physics} \textbf{\bibinfo{volume}{85}},
  \bibinfo{pages}{135} (\bibinfo{year}{2013}).

\bibitem[{\citenamefont{Chen et~al.}(2015)\citenamefont{Chen, Wang, and
  Granick}}]{chen2015memoryless}
\bibinfo{author}{\bibfnamefont{K.}~\bibnamefont{Chen}},
  \bibinfo{author}{\bibfnamefont{B.}~\bibnamefont{Wang}}, \bibnamefont{and}
  \bibinfo{author}{\bibfnamefont{S.}~\bibnamefont{Granick}},
  \bibinfo{journal}{Nature Materials} \textbf{\bibinfo{volume}{14}},
  \bibinfo{pages}{589} (\bibinfo{year}{2015}).

\bibitem[{\citenamefont{Ba et~al.}(2018)\citenamefont{Ba, Raghavan, Kiselyov,
  and Yang}}]{ba2018whole}
\bibinfo{author}{\bibfnamefont{Q.}~\bibnamefont{Ba}},
  \bibinfo{author}{\bibfnamefont{G.}~\bibnamefont{Raghavan}},
  \bibinfo{author}{\bibfnamefont{K.}~\bibnamefont{Kiselyov}}, \bibnamefont{and}
  \bibinfo{author}{\bibfnamefont{G.}~\bibnamefont{Yang}},
  \bibinfo{journal}{Cell reports} \textbf{\bibinfo{volume}{23}},
  \bibinfo{pages}{3591} (\bibinfo{year}{2018}).

\bibitem[{\citenamefont{Fedotov et~al.}(2018)\citenamefont{Fedotov, Korabel,
  Waigh, Han, and Allan}}]{fedotov2018memory}
\bibinfo{author}{\bibfnamefont{S.}~\bibnamefont{Fedotov}},
  \bibinfo{author}{\bibfnamefont{N.}~\bibnamefont{Korabel}},
  \bibinfo{author}{\bibfnamefont{T.~A.} \bibnamefont{Waigh}},
  \bibinfo{author}{\bibfnamefont{D.}~\bibnamefont{Han}}, \bibnamefont{and}
  \bibinfo{author}{\bibfnamefont{V.~J.} \bibnamefont{Allan}},
  \bibinfo{journal}{Physical Review E} \textbf{\bibinfo{volume}{98}},
  \bibinfo{pages}{042136} (\bibinfo{year}{2018}).

\bibitem[{\citenamefont{Huda et~al.}(2018)\citenamefont{Huda, Weigelin, Wolf,
  Tretiakov, Polev, Wilk, Iwasa, Emami, Narojczyk, Banaszak
  et~al.}}]{huda2018levy}
\bibinfo{author}{\bibfnamefont{S.}~\bibnamefont{Huda}},
  \bibinfo{author}{\bibfnamefont{B.}~\bibnamefont{Weigelin}},
  \bibinfo{author}{\bibfnamefont{K.}~\bibnamefont{Wolf}},
  \bibinfo{author}{\bibfnamefont{K.~V.} \bibnamefont{Tretiakov}},
  \bibinfo{author}{\bibfnamefont{K.}~\bibnamefont{Polev}},
  \bibinfo{author}{\bibfnamefont{G.}~\bibnamefont{Wilk}},
  \bibinfo{author}{\bibfnamefont{M.}~\bibnamefont{Iwasa}},
  \bibinfo{author}{\bibfnamefont{F.~S.} \bibnamefont{Emami}},
  \bibinfo{author}{\bibfnamefont{J.~W.} \bibnamefont{Narojczyk}},
  \bibinfo{author}{\bibfnamefont{M.}~\bibnamefont{Banaszak}},
  \bibnamefont{et~al.}, \bibinfo{journal}{Nature communications}
  \textbf{\bibinfo{volume}{9}}, \bibinfo{pages}{1} (\bibinfo{year}{2018}).

\bibitem[{\citenamefont{Sch{\"u}tz and Trimper}(2004)}]{schutz2004elephants}
\bibinfo{author}{\bibfnamefont{G.~M.} \bibnamefont{Sch{\"u}tz}}
  \bibnamefont{and} \bibinfo{author}{\bibfnamefont{S.}~\bibnamefont{Trimper}},
  \bibinfo{journal}{Physical Review E} \textbf{\bibinfo{volume}{70}},
  \bibinfo{pages}{045101} (\bibinfo{year}{2004}).

\bibitem[{\citenamefont{Kumar et~al.}(2010)\citenamefont{Kumar, Harbola, and
  Lindenberg}}]{kumar2010memory}
\bibinfo{author}{\bibfnamefont{N.}~\bibnamefont{Kumar}},
  \bibinfo{author}{\bibfnamefont{U.}~\bibnamefont{Harbola}}, \bibnamefont{and}
  \bibinfo{author}{\bibfnamefont{K.}~\bibnamefont{Lindenberg}},
  \bibinfo{journal}{Physical Review E} \textbf{\bibinfo{volume}{82}},
  \bibinfo{pages}{021101} (\bibinfo{year}{2010}).

\bibitem[{\citenamefont{Da~Silva et~al.}(2013)\citenamefont{Da~Silva, Cressoni,
  Sch{\"u}tz, Viswanathan, and Trimper}}]{da2013non}
\bibinfo{author}{\bibfnamefont{M.}~\bibnamefont{Da~Silva}},
  \bibinfo{author}{\bibfnamefont{J.~C.} \bibnamefont{Cressoni}},
  \bibinfo{author}{\bibfnamefont{G.~M.} \bibnamefont{Sch{\"u}tz}},
  \bibinfo{author}{\bibfnamefont{G.}~\bibnamefont{Viswanathan}},
  \bibnamefont{and} \bibinfo{author}{\bibfnamefont{S.}~\bibnamefont{Trimper}},
  \bibinfo{journal}{Physical Review E} \textbf{\bibinfo{volume}{88}},
  \bibinfo{pages}{022115} (\bibinfo{year}{2013}).

\bibitem[{\citenamefont{Budini}(2016)}]{budini2016inhomogeneous}
\bibinfo{author}{\bibfnamefont{A.~A.} \bibnamefont{Budini}},
  \bibinfo{journal}{Physical Review E} \textbf{\bibinfo{volume}{94}},
  \bibinfo{pages}{052142} (\bibinfo{year}{2016}).

\bibitem[{\citenamefont{Baur and Bertoin}(2016)}]{baur2016elephant}
\bibinfo{author}{\bibfnamefont{E.}~\bibnamefont{Baur}} \bibnamefont{and}
  \bibinfo{author}{\bibfnamefont{J.}~\bibnamefont{Bertoin}},
  \bibinfo{journal}{Physical review E} \textbf{\bibinfo{volume}{94}},
  \bibinfo{pages}{052134} (\bibinfo{year}{2016}).

\bibitem[{\citenamefont{Pemantle}(1988)}]{pemantle1988phase}
\bibinfo{author}{\bibfnamefont{R.}~\bibnamefont{Pemantle}},
  \bibinfo{journal}{The Annals of Probability} pp. \bibinfo{pages}{1229--1241}
  (\bibinfo{year}{1988}).

\bibitem[{\citenamefont{Tejedor et~al.}(2012)\citenamefont{Tejedor, Voituriez,
  and B{\'e}nichou}}]{tejedor2012optimizing}
\bibinfo{author}{\bibfnamefont{V.}~\bibnamefont{Tejedor}},
  \bibinfo{author}{\bibfnamefont{R.}~\bibnamefont{Voituriez}},
  \bibnamefont{and}
  \bibinfo{author}{\bibfnamefont{O.}~\bibnamefont{B{\'e}nichou}},
  \bibinfo{journal}{Physical review letters} \textbf{\bibinfo{volume}{108}},
  \bibinfo{pages}{088103} (\bibinfo{year}{2012}).

\bibitem[{\citenamefont{Lanoisel{\'e}e
  et~al.}(2018)\citenamefont{Lanoisel{\'e}e, Moutal, and
  Grebenkov}}]{lanoiselee2018diffusion}
\bibinfo{author}{\bibfnamefont{Y.}~\bibnamefont{Lanoisel{\'e}e}},
  \bibinfo{author}{\bibfnamefont{N.}~\bibnamefont{Moutal}}, \bibnamefont{and}
  \bibinfo{author}{\bibfnamefont{D.~S.} \bibnamefont{Grebenkov}},
  \bibinfo{journal}{Nature communications} \textbf{\bibinfo{volume}{9}},
  \bibinfo{pages}{1} (\bibinfo{year}{2018}).

\bibitem[{\citenamefont{Meyer and Rieger}(2021)}]{meyer2021optimal}
\bibinfo{author}{\bibfnamefont{H.}~\bibnamefont{Meyer}} \bibnamefont{and}
  \bibinfo{author}{\bibfnamefont{H.}~\bibnamefont{Rieger}},
  \bibinfo{journal}{Physical Review Letters} \textbf{\bibinfo{volume}{127}},
  \bibinfo{pages}{070601} (\bibinfo{year}{2021}).

\bibitem[{\citenamefont{Han et~al.}(2021{\natexlab{a}})\citenamefont{Han,
  da~Silva, Korabel, and Fedotov}}]{han2021self}
\bibinfo{author}{\bibfnamefont{D.}~\bibnamefont{Han}},
  \bibinfo{author}{\bibfnamefont{M.~A.} \bibnamefont{da~Silva}},
  \bibinfo{author}{\bibfnamefont{N.}~\bibnamefont{Korabel}}, \bibnamefont{and}
  \bibinfo{author}{\bibfnamefont{S.}~\bibnamefont{Fedotov}},
  \bibinfo{journal}{Physical Review E} \textbf{\bibinfo{volume}{103}},
  \bibinfo{pages}{022132} (\bibinfo{year}{2021}{\natexlab{a}}).

\bibitem[{\citenamefont{Fedotov et~al.}(2022)\citenamefont{Fedotov, Han,
  Ivanov, and da~Silva}}]{fedotov2022superdiffusion}
\bibinfo{author}{\bibfnamefont{S.}~\bibnamefont{Fedotov}},
  \bibinfo{author}{\bibfnamefont{D.}~\bibnamefont{Han}},
  \bibinfo{author}{\bibfnamefont{A.~O.} \bibnamefont{Ivanov}},
  \bibnamefont{and} \bibinfo{author}{\bibfnamefont{M.~A.}
  \bibnamefont{da~Silva}}, \bibinfo{journal}{Physical Review E}
  \textbf{\bibinfo{volume}{105}}, \bibinfo{pages}{014126}
  (\bibinfo{year}{2022}).

\bibitem[{\citenamefont{Han et~al.}(2021{\natexlab{b}})\citenamefont{Han,
  Alexandrov, Gavrilova, and Fedotov}}]{han2021anomalous}
\bibinfo{author}{\bibfnamefont{D.}~\bibnamefont{Han}},
  \bibinfo{author}{\bibfnamefont{D.~V.} \bibnamefont{Alexandrov}},
  \bibinfo{author}{\bibfnamefont{A.}~\bibnamefont{Gavrilova}},
  \bibnamefont{and} \bibinfo{author}{\bibfnamefont{S.}~\bibnamefont{Fedotov}},
  \bibinfo{journal}{Fractal and Fractional} \textbf{\bibinfo{volume}{5}},
  \bibinfo{pages}{221} (\bibinfo{year}{2021}{\natexlab{b}}).

\bibitem[{\citenamefont{Ariel et~al.}(2015)\citenamefont{Ariel, Rabani,
  Benisty, Partridge, Harshey, and Be'Er}}]{ariel2015swarming}
\bibinfo{author}{\bibfnamefont{G.}~\bibnamefont{Ariel}},
  \bibinfo{author}{\bibfnamefont{A.}~\bibnamefont{Rabani}},
  \bibinfo{author}{\bibfnamefont{S.}~\bibnamefont{Benisty}},
  \bibinfo{author}{\bibfnamefont{J.~D.} \bibnamefont{Partridge}},
  \bibinfo{author}{\bibfnamefont{R.~M.} \bibnamefont{Harshey}},
  \bibnamefont{and} \bibinfo{author}{\bibfnamefont{A.}~\bibnamefont{Be'Er}},
  \bibinfo{journal}{Nature Communications} \textbf{\bibinfo{volume}{6}},
  \bibinfo{pages}{1} (\bibinfo{year}{2015}).

\bibitem[{\citenamefont{Han et~al.}(2020)\citenamefont{Han, Korabel, Chen,
  Johnston, Gavrilova, Allan, Fedotov, and Waigh}}]{han2020deciphering}
\bibinfo{author}{\bibfnamefont{D.}~\bibnamefont{Han}},
  \bibinfo{author}{\bibfnamefont{N.}~\bibnamefont{Korabel}},
  \bibinfo{author}{\bibfnamefont{R.}~\bibnamefont{Chen}},
  \bibinfo{author}{\bibfnamefont{M.}~\bibnamefont{Johnston}},
  \bibinfo{author}{\bibfnamefont{A.}~\bibnamefont{Gavrilova}},
  \bibinfo{author}{\bibfnamefont{V.~J.} \bibnamefont{Allan}},
  \bibinfo{author}{\bibfnamefont{S.}~\bibnamefont{Fedotov}}, \bibnamefont{and}
  \bibinfo{author}{\bibfnamefont{T.~A.} \bibnamefont{Waigh}},
  \bibinfo{journal}{Elife} \textbf{\bibinfo{volume}{9}} (\bibinfo{year}{2020}).

\bibitem[{\citenamefont{Altschuler and Wu}(2010)}]{altschuler2010cellular}
\bibinfo{author}{\bibfnamefont{S.~J.} \bibnamefont{Altschuler}}
  \bibnamefont{and} \bibinfo{author}{\bibfnamefont{L.~F.} \bibnamefont{Wu}},
  \bibinfo{journal}{Cell} \textbf{\bibinfo{volume}{141}}, \bibinfo{pages}{559}
  (\bibinfo{year}{2010}).

\bibitem[{\citenamefont{Muotri et~al.}(2005)\citenamefont{Muotri, Chu,
  Marchetto, Deng, Moran, and Gage}}]{muotri2005somatic}
\bibinfo{author}{\bibfnamefont{A.~R.} \bibnamefont{Muotri}},
  \bibinfo{author}{\bibfnamefont{V.~T.} \bibnamefont{Chu}},
  \bibinfo{author}{\bibfnamefont{M.~C.} \bibnamefont{Marchetto}},
  \bibinfo{author}{\bibfnamefont{W.}~\bibnamefont{Deng}},
  \bibinfo{author}{\bibfnamefont{J.~V.} \bibnamefont{Moran}}, \bibnamefont{and}
  \bibinfo{author}{\bibfnamefont{F.~H.} \bibnamefont{Gage}},
  \bibinfo{journal}{nature} \textbf{\bibinfo{volume}{435}},
  \bibinfo{pages}{903} (\bibinfo{year}{2005}).

\bibitem[{\citenamefont{Lomvardas et~al.}(2006)\citenamefont{Lomvardas, Barnea,
  Pisapia, Mendelsohn, Kirkland, and Axel}}]{lomvardas2006interchromosomal}
\bibinfo{author}{\bibfnamefont{S.}~\bibnamefont{Lomvardas}},
  \bibinfo{author}{\bibfnamefont{G.}~\bibnamefont{Barnea}},
  \bibinfo{author}{\bibfnamefont{D.~J.} \bibnamefont{Pisapia}},
  \bibinfo{author}{\bibfnamefont{M.}~\bibnamefont{Mendelsohn}},
  \bibinfo{author}{\bibfnamefont{J.}~\bibnamefont{Kirkland}}, \bibnamefont{and}
  \bibinfo{author}{\bibfnamefont{R.}~\bibnamefont{Axel}},
  \bibinfo{journal}{Cell} \textbf{\bibinfo{volume}{126}}, \bibinfo{pages}{403}
  (\bibinfo{year}{2006}).

\bibitem[{\citenamefont{Coufal et~al.}(2009)\citenamefont{Coufal, Garcia-Perez,
  Peng, Yeo, Mu, Lovci, Morell, O’Shea, Moran, and Gage}}]{coufal2009l1}
\bibinfo{author}{\bibfnamefont{N.~G.} \bibnamefont{Coufal}},
  \bibinfo{author}{\bibfnamefont{J.~L.} \bibnamefont{Garcia-Perez}},
  \bibinfo{author}{\bibfnamefont{G.~E.} \bibnamefont{Peng}},
  \bibinfo{author}{\bibfnamefont{G.~W.} \bibnamefont{Yeo}},
  \bibinfo{author}{\bibfnamefont{Y.}~\bibnamefont{Mu}},
  \bibinfo{author}{\bibfnamefont{M.~T.} \bibnamefont{Lovci}},
  \bibinfo{author}{\bibfnamefont{M.}~\bibnamefont{Morell}},
  \bibinfo{author}{\bibfnamefont{K.~S.} \bibnamefont{O’Shea}},
  \bibinfo{author}{\bibfnamefont{J.~V.} \bibnamefont{Moran}}, \bibnamefont{and}
  \bibinfo{author}{\bibfnamefont{F.~H.} \bibnamefont{Gage}},
  \bibinfo{journal}{Nature} \textbf{\bibinfo{volume}{460}},
  \bibinfo{pages}{1127} (\bibinfo{year}{2009}).

\bibitem[{\citenamefont{Zhu et~al.}(2021)\citenamefont{Zhu, Li, Liao, Yin,
  Chen, Wang, Dai, Yi, Ge, Miao et~al.}}]{zhu2021metabolomic}
\bibinfo{author}{\bibfnamefont{H.}~\bibnamefont{Zhu}},
  \bibinfo{author}{\bibfnamefont{Q.}~\bibnamefont{Li}},
  \bibinfo{author}{\bibfnamefont{T.}~\bibnamefont{Liao}},
  \bibinfo{author}{\bibfnamefont{X.}~\bibnamefont{Yin}},
  \bibinfo{author}{\bibfnamefont{Q.}~\bibnamefont{Chen}},
  \bibinfo{author}{\bibfnamefont{Z.}~\bibnamefont{Wang}},
  \bibinfo{author}{\bibfnamefont{M.}~\bibnamefont{Dai}},
  \bibinfo{author}{\bibfnamefont{L.}~\bibnamefont{Yi}},
  \bibinfo{author}{\bibfnamefont{S.}~\bibnamefont{Ge}},
  \bibinfo{author}{\bibfnamefont{C.}~\bibnamefont{Miao}}, \bibnamefont{et~al.},
  \bibinfo{journal}{Nature Methods} \textbf{\bibinfo{volume}{18}},
  \bibinfo{pages}{788} (\bibinfo{year}{2021}).

\bibitem[{\citenamefont{Fedotov et~al.}(2021)\citenamefont{Fedotov, Han,
  Zubarev, Johnston, and Allan}}]{fedotov2021variable}
\bibinfo{author}{\bibfnamefont{S.}~\bibnamefont{Fedotov}},
  \bibinfo{author}{\bibfnamefont{D.}~\bibnamefont{Han}},
  \bibinfo{author}{\bibfnamefont{A.~Y.} \bibnamefont{Zubarev}},
  \bibinfo{author}{\bibfnamefont{M.}~\bibnamefont{Johnston}}, \bibnamefont{and}
  \bibinfo{author}{\bibfnamefont{V.~J.} \bibnamefont{Allan}},
  \bibinfo{journal}{Philosophical Transactions of the Royal Society A}
  \textbf{\bibinfo{volume}{379}}, \bibinfo{pages}{20200317}
  (\bibinfo{year}{2021}).

\bibitem[{\citenamefont{Fox et~al.}(2021)\citenamefont{Fox, Barkai, and
  Krapf}}]{fox2021aging}
\bibinfo{author}{\bibfnamefont{Z.~R.} \bibnamefont{Fox}},
  \bibinfo{author}{\bibfnamefont{E.}~\bibnamefont{Barkai}}, \bibnamefont{and}
  \bibinfo{author}{\bibfnamefont{D.}~\bibnamefont{Krapf}},
  \bibinfo{journal}{Nature communications} \textbf{\bibinfo{volume}{12}},
  \bibinfo{pages}{1} (\bibinfo{year}{2021}).

\bibitem[{\citenamefont{Hughes}(1995)}]{hughes1995random}
\bibinfo{author}{\bibfnamefont{B.}~\bibnamefont{Hughes}},
  \emph{\bibinfo{title}{Random walks and random environments, volume 1: Random
  walks clarendon}} (\bibinfo{year}{1995}).

\bibitem[{\citenamefont{Hughes}(1996)}]{hughes1996random}
\bibinfo{author}{\bibfnamefont{B.}~\bibnamefont{Hughes}},
  \emph{\bibinfo{title}{Random walks and random environments, volume 2:random
  environments}} (\bibinfo{year}{1996}).

\bibitem[{\citenamefont{Fedotov and Han}(2019)}]{fedotov2019asymptotic}
\bibinfo{author}{\bibfnamefont{S.}~\bibnamefont{Fedotov}} \bibnamefont{and}
  \bibinfo{author}{\bibfnamefont{D.}~\bibnamefont{Han}},
  \bibinfo{journal}{Physical review letters} \textbf{\bibinfo{volume}{123}},
  \bibinfo{pages}{050602} (\bibinfo{year}{2019}).

\bibitem[{\citenamefont{Flores-Rodriguez
  et~al.}(2011)\citenamefont{Flores-Rodriguez, Rogers, Kenwright, Waigh,
  Woodman, and Allan}}]{flores2011roles}
\bibinfo{author}{\bibfnamefont{N.}~\bibnamefont{Flores-Rodriguez}},
  \bibinfo{author}{\bibfnamefont{S.~S.} \bibnamefont{Rogers}},
  \bibinfo{author}{\bibfnamefont{D.~A.} \bibnamefont{Kenwright}},
  \bibinfo{author}{\bibfnamefont{T.~A.} \bibnamefont{Waigh}},
  \bibinfo{author}{\bibfnamefont{P.~G.} \bibnamefont{Woodman}},
  \bibnamefont{and} \bibinfo{author}{\bibfnamefont{V.~J.} \bibnamefont{Allan}},
  \bibinfo{journal}{PloS one} \textbf{\bibinfo{volume}{6}},
  \bibinfo{pages}{e24479} (\bibinfo{year}{2011}).

\bibitem[{\citenamefont{Wang et~al.}(2012)\citenamefont{Wang, Kuo, Bae, and
  Granick}}]{wang2012brownian}
\bibinfo{author}{\bibfnamefont{B.}~\bibnamefont{Wang}},
  \bibinfo{author}{\bibfnamefont{J.}~\bibnamefont{Kuo}},
  \bibinfo{author}{\bibfnamefont{S.~C.} \bibnamefont{Bae}}, \bibnamefont{and}
  \bibinfo{author}{\bibfnamefont{S.}~\bibnamefont{Granick}},
  \bibinfo{journal}{Nature materials} \textbf{\bibinfo{volume}{11}},
  \bibinfo{pages}{481} (\bibinfo{year}{2012}).

\bibitem[{\citenamefont{Chubynsky and Slater}(2014)}]{chubynsky2014diffusing}
\bibinfo{author}{\bibfnamefont{M.~V.} \bibnamefont{Chubynsky}}
  \bibnamefont{and} \bibinfo{author}{\bibfnamefont{G.~W.}
  \bibnamefont{Slater}}, \bibinfo{journal}{Physical review letters}
  \textbf{\bibinfo{volume}{113}}, \bibinfo{pages}{098302}
  (\bibinfo{year}{2014}).

\bibitem[{\citenamefont{Jain and Sebastian}(2016)}]{jain2016diffusing}
\bibinfo{author}{\bibfnamefont{R.}~\bibnamefont{Jain}} \bibnamefont{and}
  \bibinfo{author}{\bibfnamefont{K.~L.} \bibnamefont{Sebastian}},
  \bibinfo{journal}{The Journal of Physical Chemistry B}
  \textbf{\bibinfo{volume}{120}}, \bibinfo{pages}{9215} (\bibinfo{year}{2016}).

\bibitem[{\citenamefont{Metzler}(2020)}]{metzler2020superstatistics}
\bibinfo{author}{\bibfnamefont{R.}~\bibnamefont{Metzler}},
  \bibinfo{journal}{The European Physical Journal Special Topics}
  \textbf{\bibinfo{volume}{229}}, \bibinfo{pages}{711} (\bibinfo{year}{2020}).

\bibitem[{\citenamefont{Grebenkov et~al.}(2021)\citenamefont{Grebenkov,
  Sposini, Metzler, Oshanin, and Seno}}]{grebenkov2021exact}
\bibinfo{author}{\bibfnamefont{D.~S.} \bibnamefont{Grebenkov}},
  \bibinfo{author}{\bibfnamefont{V.}~\bibnamefont{Sposini}},
  \bibinfo{author}{\bibfnamefont{R.}~\bibnamefont{Metzler}},
  \bibinfo{author}{\bibfnamefont{G.}~\bibnamefont{Oshanin}}, \bibnamefont{and}
  \bibinfo{author}{\bibfnamefont{F.}~\bibnamefont{Seno}}, \bibinfo{journal}{New
  Journal of Physics} \textbf{\bibinfo{volume}{23}}, \bibinfo{pages}{023014}
  (\bibinfo{year}{2021}).

\bibitem[{\citenamefont{Sandev et~al.}(2022)\citenamefont{Sandev, Domazetoski,
  Kocarev, Metzler, and Chechkin}}]{sandev2022heterogeneous}
\bibinfo{author}{\bibfnamefont{T.}~\bibnamefont{Sandev}},
  \bibinfo{author}{\bibfnamefont{V.}~\bibnamefont{Domazetoski}},
  \bibinfo{author}{\bibfnamefont{L.}~\bibnamefont{Kocarev}},
  \bibinfo{author}{\bibfnamefont{R.}~\bibnamefont{Metzler}}, \bibnamefont{and}
  \bibinfo{author}{\bibfnamefont{A.}~\bibnamefont{Chechkin}},
  \bibinfo{journal}{Journal of Physics A: Mathematical and Theoretical}
  \textbf{\bibinfo{volume}{55}}, \bibinfo{pages}{074003}
  (\bibinfo{year}{2022}).

\bibitem[{\citenamefont{Chechkin et~al.}(2005)\citenamefont{Chechkin, Gorenflo,
  and Sokolov}}]{chechkin2005fractional}
\bibinfo{author}{\bibfnamefont{A.~V.} \bibnamefont{Chechkin}},
  \bibinfo{author}{\bibfnamefont{R.}~\bibnamefont{Gorenflo}}, \bibnamefont{and}
  \bibinfo{author}{\bibfnamefont{I.~M.} \bibnamefont{Sokolov}},
  \bibinfo{journal}{Journal of Physics A: Mathematical and General}
  \textbf{\bibinfo{volume}{38}}, \bibinfo{pages}{L679} (\bibinfo{year}{2005}).

\bibitem[{\citenamefont{Korabel and Barkai}(2010)}]{korabel2010paradoxes}
\bibinfo{author}{\bibfnamefont{N.}~\bibnamefont{Korabel}} \bibnamefont{and}
  \bibinfo{author}{\bibfnamefont{E.}~\bibnamefont{Barkai}},
  \bibinfo{journal}{Physical review letters} \textbf{\bibinfo{volume}{104}},
  \bibinfo{pages}{170603} (\bibinfo{year}{2010}).

\bibitem[{\citenamefont{Berry and Soula}(2014)}]{berry2014spatial}
\bibinfo{author}{\bibfnamefont{H.}~\bibnamefont{Berry}} \bibnamefont{and}
  \bibinfo{author}{\bibfnamefont{H.~A.} \bibnamefont{Soula}},
  \bibinfo{journal}{Frontiers in physiology} \textbf{\bibinfo{volume}{5}},
  \bibinfo{pages}{437} (\bibinfo{year}{2014}).

\bibitem[{\citenamefont{Heinsalu et~al.}(2007)\citenamefont{Heinsalu,
  Patriarca, Goychuk, and H{\"a}nggi}}]{heinsalu2007use}
\bibinfo{author}{\bibfnamefont{E.}~\bibnamefont{Heinsalu}},
  \bibinfo{author}{\bibfnamefont{M.}~\bibnamefont{Patriarca}},
  \bibinfo{author}{\bibfnamefont{I.}~\bibnamefont{Goychuk}}, \bibnamefont{and}
  \bibinfo{author}{\bibfnamefont{P.}~\bibnamefont{H{\"a}nggi}},
  \bibinfo{journal}{Physical review letters} \textbf{\bibinfo{volume}{99}},
  \bibinfo{pages}{120602} (\bibinfo{year}{2007}).

\bibitem[{\citenamefont{Feller}(1971)}]{feller1971introduction}
\bibinfo{author}{\bibfnamefont{W.}~\bibnamefont{Feller}},
  \emph{\bibinfo{title}{An introduction to probability theory and its
  applications}}, vol.~\bibinfo{volume}{1} (\bibinfo{year}{1971}).

\bibitem[{\citenamefont{Laskin}(2003)}]{laskin2003fractional}
\bibinfo{author}{\bibfnamefont{N.}~\bibnamefont{Laskin}},
  \bibinfo{journal}{Communications in Nonlinear Science and Numerical
  Simulation} \textbf{\bibinfo{volume}{8}}, \bibinfo{pages}{201}
  (\bibinfo{year}{2003}).

\bibitem[{\citenamefont{Fulger et~al.}(2008)\citenamefont{Fulger, Scalas, and
  Germano}}]{fulger2008monte}
\bibinfo{author}{\bibfnamefont{D.}~\bibnamefont{Fulger}},
  \bibinfo{author}{\bibfnamefont{E.}~\bibnamefont{Scalas}}, \bibnamefont{and}
  \bibinfo{author}{\bibfnamefont{G.}~\bibnamefont{Germano}},
  \bibinfo{journal}{Physical Review E} \textbf{\bibinfo{volume}{77}},
  \bibinfo{pages}{021122} (\bibinfo{year}{2008}).

\bibitem[{\citenamefont{Kozubowski and
  Rachev}(1999)}]{kozubowski1999univariate}
\bibinfo{author}{\bibfnamefont{T.~J.} \bibnamefont{Kozubowski}}
  \bibnamefont{and} \bibinfo{author}{\bibfnamefont{S.~T.}
  \bibnamefont{Rachev}}, \bibinfo{journal}{Journal of Computational Analysis
  and Applications} \textbf{\bibinfo{volume}{1}}, \bibinfo{pages}{177}
  (\bibinfo{year}{1999}).

\bibitem[{\citenamefont{Cox and Miller}(1965)}]{coxmillertextbook}
\bibinfo{author}{\bibfnamefont{D.~R.} \bibnamefont{Cox}} \bibnamefont{and}
  \bibinfo{author}{\bibfnamefont{H.~D.} \bibnamefont{Miller}},
  \emph{\bibinfo{title}{The theory of stochastic processes}}
  (\bibinfo{publisher}{Chapman and Hall}, \bibinfo{year}{1965}).

\bibitem[{\citenamefont{Metzler et~al.}(1998)\citenamefont{Metzler, Klafter,
  and Sokolov}}]{metzler1998anomalous}
\bibinfo{author}{\bibfnamefont{R.}~\bibnamefont{Metzler}},
  \bibinfo{author}{\bibfnamefont{J.}~\bibnamefont{Klafter}}, \bibnamefont{and}
  \bibinfo{author}{\bibfnamefont{I.~M.} \bibnamefont{Sokolov}},
  \bibinfo{journal}{Physical Review E} \textbf{\bibinfo{volume}{58}},
  \bibinfo{pages}{1621} (\bibinfo{year}{1998}).

\bibitem[{\citenamefont{Goychuk et~al.}(2006)\citenamefont{Goychuk, Heinsalu,
  Patriarca, Schmid, and H{\"a}nggi}}]{goychuk2006current}
\bibinfo{author}{\bibfnamefont{I.}~\bibnamefont{Goychuk}},
  \bibinfo{author}{\bibfnamefont{E.}~\bibnamefont{Heinsalu}},
  \bibinfo{author}{\bibfnamefont{M.}~\bibnamefont{Patriarca}},
  \bibinfo{author}{\bibfnamefont{G.}~\bibnamefont{Schmid}}, \bibnamefont{and}
  \bibinfo{author}{\bibfnamefont{P.}~\bibnamefont{H{\"a}nggi}},
  \bibinfo{journal}{Physical Review E} \textbf{\bibinfo{volume}{73}},
  \bibinfo{pages}{020101} (\bibinfo{year}{2006}).

\bibitem[{\citenamefont{Shlesinger}(1974)}]{shlesinger1974asymptotic}
\bibinfo{author}{\bibfnamefont{M.~F.} \bibnamefont{Shlesinger}},
  \bibinfo{journal}{Journal of Statistical Physics}
  \textbf{\bibinfo{volume}{10}}, \bibinfo{pages}{421} (\bibinfo{year}{1974}).

\bibitem[{\citenamefont{Schuss et~al.}(2019)\citenamefont{Schuss, Basnayake,
  and Holcman}}]{schuss2019redundancy}
\bibinfo{author}{\bibfnamefont{Z.}~\bibnamefont{Schuss}},
  \bibinfo{author}{\bibfnamefont{K.}~\bibnamefont{Basnayake}},
  \bibnamefont{and} \bibinfo{author}{\bibfnamefont{D.}~\bibnamefont{Holcman}},
  \bibinfo{journal}{Physics of life reviews} \textbf{\bibinfo{volume}{28}},
  \bibinfo{pages}{52} (\bibinfo{year}{2019}).

\end{thebibliography}
	
	\begin{acknowledgments}
		\textbf{S.F. acknowledges funding from EPSRC grant no. EP/V008641/1.
		D.H. acknowledges funding from the MRC Laboratory of Molecular Biology.}
		
		The authors would like to thank the Isaac Newton Institute for Mathematical Sciences, Cambridge, for support and hospitality during the programme ``Fractional Differential Equations [FDE2]'' where work on this paper was undertaken. This work was supported by EPSRC grant no. EP/R014604/1.
	\end{acknowledgments}
	
\end{document}